# Stormwater on the Margins: Influence of Race, Gender, and Education on Willingness to Participate in Stormwater Management

Rachel D. Scarlett[a,c*], Mangala Subramaniam[b], Sara K. McMillan[a,c], Anastasia T. Ingermann[c], Sandra M. Clinton[d]

[a]Ecological Sciences and Engineering Interdisciplinary Program, Purdue University, West Lafayette, IN, USA; [b]Susan Bulkeley Butler Center for Leadership Excellence, Purdue University, West Lafayette, IN, USA; [c]Department of Agricultural and Biological Engineering, Purdue University, West Lafayette, IN, USA; [d]Department of Geography and Earth Sciences, University of North Carolina at Charlotte, Charlotte, NC, USA

*Corresponding author email address: rdscarlett22@gmail.com

**Abstract**

Stormwater has immense impacts on urban flooding and water quality, leaving the marginalized and the impoverished disproportionately impacted by and vulnerable to stormwater hazards. However, the environmental health concerns of socially and economically marginalized individuals are largely underestimated. Through regression analysis of data from three longitudinal surveys, this article examines if and how an individual's race, gender, and education level help predict one's concern about and willingness to participate in stormwater management. We found that people of color, women, and less-educated respondents had a greater willingness to participate in stormwater management than White, male, and more-educated respondents, and their concern about local stormwater hazards drove their willingness to participate. Our analysis suggests that physical exposure and high vulnerability to stormwater hazards may shape an individual's concern about and willingness to participate in stormwater management.

# 1. Introduction

Urban stormwater has drawn water managers' attention because of its deleterious impacts on flooding and water quality in surrounding streams, rivers, lakes, and coastal zones (e.g., Meyer et al., 2005; O'Driscoll et al., 2010). Additionally, nuisance algal blooms and increased sediment loads can threaten drinking water reservoirs (Carmichael and Boyer, 2016; Gaffield et al., 2003). The magnitude of this problem is underscored as millions of people experience flood-related damage yearly, and the cost of mitigating stormwater externalities escalates (Brody et al., 2007). While flood and water quality hazards are a serious threat to urban communities worldwide, notably, these hazards disproportionately affect socially and economically marginalized communities. Women, the impoverished, and racially marginalized individuals are at the highest risk of flooding and impaired water quality and often have the highest barriers to recovery from emergencies (Enarson and Fordham, 2000; Liévanos, 2017; Qiang, 2019). Such an inequitable distribution of environmental degradation triggers broad social concerns about flooding, water quality, and ecological integrity and creates a unique socioecological problem that requires solutions across social and technical viewpoints.

While social inequalities are persistent in urban water systems, conventional management of stormwater is technocratic— centered on engineering strategies that convey water, sediment, and nutrients out of sight (Finewood, 2016). Technocratic governance hides stormwater's socioecological complexity, reinforces the public perception that stormwater governance is expert-driven, and, ultimately, isolates stormwater management from the public (Dhakal and Chevalier, 2016). The focus on technological solutions ignores the structural and institutional drivers of inequitable impacts and ultimately can perpetuate inequality throughout the socioecological system. Municipalities need broader approaches to stormwater management that engage communities across socioeconomic backgrounds— approaches that will improve access to stormwater management services and address the growing threats of climate change, urban growth, and socioeconomic inequality.

Scholars have shown that individuals' social and economic status is an important predictor of their concern about and participation in environmental management broadly, but stormwater management has been overlooked. A recent study examining a broad range of environmental concerns illustrates that diverse segments of the American public underestimated the environmental concerns of racially marginalized and impoverished individuals (Pearson et al., 2018). Despite public perception, scholars have illustrated that people of color, the impoverished, and women tend to be just as or more concerned about environmental issues than more socioeconomically privileged groups, especially about issues related to environmental racism and risk exposure (Lazri and Konisky, 2019; Macias, 2016a). Stormwater hazards align with traditional environmental racism issues, like toxic waste, regarding the inequitable distribution of vulnerability and outcomes (Debbage, 2019). Yet, notably, the interplay between inequitable experiences and technical decision-making and knowledge presents an additional complexity in stormwater management that is not completely understood. Daily experiences of stormwater hazards can raise public awareness of stormwater problems; alternatively, inaccessible technical knowledge and management can lead to the perception that stormwater is not a social and environmental problem and certainly not one that engages the public.

Using secondary data from a survey of individuals conducted in Charlotte, North Carolina, USA, we examined whether and how individuals' race, gender, and education level help predict their willingness to participate in stormwater management. Additionally, we investigated how these patterns change based on different forms of participation in stormwater management, including individuals' willingness to volunteer for stream cleanups and willingness to pay more in stormwater fees.

## 2. Literature Review

### 2.1. Social Marginalization and Environmental Concern

Numerous empirical studies have shown that a person's race, class, and gender are important predictors of their environmental concern. Environmental concern is a broad construct that is often conceptualized as a general attitude towards environmental protection. More recently, sense of environmental risk has been included as a key facet of environmental concern in recognition of the direct influence of environmental threats on individuals' attitudes towards the environment (Mohai and Bryant 1998; Macias, 2016a).

Early literature on this topic has suggested that Black people are less concerned about environmental degradation than White people (Hershey and Hill, 1977; Hohm, 1976; Kreger, 1973). For instance, Hohm (1976) conducted a survey on the relationship between one's race and concern for air pollution and found that White respondents had a higher perception of the severity of air pollution and related health risks than Black respondents. The author's explanation relied on Maslow's hierarchy of needs theory, which supports the claim that economically disadvantaged groups—assumed to be the case for Black respondents— lack concern for the environment because they focus on fundamental needs like food, housing security, and healthcare (Maslow, 1954). Critics, however, have argued that Maslow's hierarchy of needs fails to recognize the dependence of basic needs on environmental conditions (Mohai and Bryant, 1998). Water pollution and stormwater flooding can threaten one's housing security and access to safe drinking water. Others find a lack of support for Maslow's hierarchy of needs because race is a significant determinant of environmental concern regardless of socioeconomic status (Hershey and Hill, 1977). With regard to a survey of young adults on concern for litter, land preservation, and endangered species protection, Hershey and Hill (1977) instead argued that White youth are more concerned about environmental pollution than Black youth due to disparate subcultural norms. At the time of the study, the mainstream environmental movement advocated economic downscaling, which seemed to threaten economic advancement goals in the Black community. Researchers suggested that environmental support in Black communities would decline during economic downturns because the economy would be prioritized. However, Jones and Carter (1994) challenged this claim by showing that Black and White people equally supported higher national spending on environmental protection throughout the 1970s and 1980s, and this support was unaffected by economic downturns.

    A more recent wave of literature has challenged the conclusion that racially marginalized people are less concerned about the environment than White people. Empirical studies began to illustrate that Black people were just as or more concerned about the environment than their wealthy and White counterparts (Caron, 1989; Jones, 1998; Mohai and Bryant, 1998). Through an empirical analysis, Mohai and Bryant (1998) investigated three theories that could explain an environmental concern gap between Black and White Americans. The first is the environmental deprivation theory—communities of color experience greater environmental burden, which subsequently increases their environmental concern. The second hypothesis is hierarchy of needs, and the third considers cultural differences between Black and White people. Cultural differences refer to disparate sociocultural experiences of nature—for example, the Black community's (assumed) negative environmental attitudes are conditioned on their lack of access

to natural spaces like national forests and beaches due to racial segregation (Finney, 2014; Taylor, 1989). Mohai and Bryant (1998) tested whether these theories applied to a range of environmental concerns, including nature preservation, global warming, plastic waste, and air pollution. Their findings did not support the hierarchy of needs hypothesis or the cultural difference hypothesis because African Americans and low-income respondents were equally concerned as White and wealthy respondents about most environmental issues. Rather, their findings support the environmental deprivation hypothesis. They found that Black and White people are similarly concerned about most environmental issues, but in reference to issues that disproportionately affect Black populations, like industrial waste, Black people's proximity to these problems drove their heightened concern.

    Some scholars have explored the influence of environmental injustices on concern about the environment (Jones and Rainey, 2006; Lazri and Konisky, 2019). Environmental justice activists and scholars have not only revealed that people of color and those with low-income are disproportionately burdened by environmental degradation but also stressed that structural forms of racism, classism, and sexism create and sustain inequitable patterns (Arp and Boeckelman, 1997; Bullard, 2008; Hines, 2001). Moreover, environmental degradation often reflects legacies of structural racism that uniquely advantage White and wealthy people, such as housing discrimination and historical redlining practices (Pulido, 2000). Jones and Rainey (2006) explored the impact of feelings of environmental injustice on environmental concern. Their findings support previous empirical studies that illustrate a heightened environmental concern in Black respondents compared to White respondents. Furthermore, they illustrate that feelings of environmental injustice drive concern: residents who felt that they were unfairly exposed to detrimental environmental conditions were more concerned. Such experiences of poor

environmental conditions in communal settings can cause people to be more conscious of environmental injustices and subsequently participate in reporting and challenging these injustices (Young and Subramaniam, 2017).

While most empirical studies focus on race, a gender dimension reveals patterns at the intersection of gender and race. Researchers have quantified that women tend to have greater environmental concern than men (Blocker and Eckberg, 1997; Chakraborty et al., 2017; Gifford and Nilsson, 2014; Tikka et al., 2000; Uyeki and Holland, 2000). In an investigation of race and gender, Kalof et al. (2000) found that White respondents reported significantly lower environmental values and beliefs than Black and Latino respondents, but they also found that gender differences in the National Environmental Paradigm scale (NEP; a measure of general environmental attitudes) only existed for White respondents. White women scored significantly higher on the NEP scale than White men. Others have found a specific "White male effect" (Brent, 2004) on environmental concern and attribute this to White men's perception that their environmental risk is low and their institutional support is high, which reduces their concern about environmental protection and risk.

The most recent wave of research extends the literature to include multiple racial and ethnic groups in nationally representative samples. Scholars have found that people of color (including multi-ethnic Latino, Asian, and African Americans) tend to be more concerned than White people about environmental issues related to environmental risks (Macias, 2016a) and environmental racism (Lazri and Konisky, 2019), such as air and water pollution. However, outside of environmental risk, people of color show similar or greater concern for locally and globally relevant environmental issues compared to White individuals (Lazri and Konisky, 2019). These studies also show that higher income (Lazri and Konisky, 2019; Macias, 2016a) and education levels (Lazri and Konisky, 2019) are correlated with lower environmental concern, and women show higher levels of environmental concern than men (Macias, 2016b). Even when controlling for other socioeconomic factors, race is still a significant predictor of environmental concern (Macias, 2016a), highlighting the interconnected but differentiated effects of marginalization based on race, gender, and class.

Few studies support the claim that racially marginalized and low-income individuals are less concerned about the environment; however, this perception is popular in the American public (Pearson et al., 2018). This public perception likely stems from conflation of environmental concern and perceived participation in the environmental movement. The mainstream environmental movement is largely White and upper class (Taylor, 2015), leading to the inference that environmental values and concerns are also White and upper class. Distortion of environmental interest in marginalized communities largely undermines the growing popularity of the environmental justice movement and places these communities in positions where stakeholders assume their disinterest in environmental governance (Finewood, 2016). Moreover, these assumptions can substantially derail coalition building and equitable decision-making.

## 2.2. From Concern to Participatory Intentions

Are differences in concern extended to participation? Substantially fewer studies have examined which individuals, based on gender, race, and education level, are willing to participate in environmentally conscious ways. Attitudes and behaviors exhibit a tenuous relationship, and research to date is inconclusive regarding the influence of social marginalization on environmental behaviors. Scholars have discussed both participation and willingness to participate in environmentally conscious behaviors. Here we conceptualize participation as actions that individuals take to improve environmental quality or mitigate environmental problems and willingness to participate as an intention to perform these behaviors. Those who express the intention to perform environmentally conscious behaviors are more likely than others to actually perform those behaviors (Hines et al., 1987).

Content validity on measures of participation and willingness to participate in environmentally conscious behaviors has been a consistent issue. Environmentally conscious behaviors frequently examined in the literature often tap into underlying issues of disproportionate economic opportunity as well as social and physical resources. For example, the frequency of recycling is often higher with individuals who are White, earn a higher income, and have a higher education level (Johnson et al., 2004; Macias, 2016a). However, Laidley (2013) determined that a significant predictor of recycling behavior is access to curbside recycling programs highlighting accessibility as an underlying issue. Likewise, individuals with higher income and education levels are often more willing to pay for environmental management (Chui and Ngai, 2016; Macias, 2016a; Newburn and Alberini, 2016). Other measures, such as purchasing chemical-free products, organic foods, and electric vehicles, show similar trends with an individual's income and education level (Laidley, 2013; Macias, 2016b). These high-status consumptive behaviors are strongly tied to social class and can just as easily align with attitudes of class distinction as they do with environmental concerns (Kennedy and Givens, 2019).

Much of the research on race and environmentally conscious behavior has shown that historically marginalized individuals participate less in environmentally conscious behaviors; however, the limitations of current measures do not capture the complexity of these behaviors. A literature review illustrates that Black and Latino people show high concern for national parks and natural resources, but they are highly underrepresented in outdoor recreational activities in parks (Roberts and Rodriguez, 2008). Likewise, marginalized individuals are grossly underrepresented in mainstream environmental groups, with most members and leaders being White and upper class (Taylor, 2014). A more nuanced look shows that Black and Latino individuals tend to lack awareness of recreational opportunities in national parks and perceive these spaces as unwelcoming and discriminatory towards communities of color (Roberts and Rodriguez, 2008). Similarly, exclusionary practices within mainstream environmental groups and failures to address the needs of working-class communities of color deter participation in mainstream environmentalism (Clarke and Agyeman, 2011; Hoover, 2017). Ultimately, measures of participation can capture broader constructs than intended, which can lead to significant biases.

Given the scant body of work on this topic and the complexity of measuring environmentally conscious behaviors and intentions to participate in those behaviors, we measure intention to participate in environmentally conscious behaviors in two ways: willingness to pay and willingness to volunteer. While willingness to pay taps into issues of economic opportunity, willingness to volunteer provides an alternate measure of participation that does not require economic investment. Hands-on and practical stewardship activities help communities closely relate to their local environments but are often ignored in discussions of public participation (but *see* Ando et al., 2020). Such actions broaden the scope of what we imagine as participation in environmental management (Eden and Bear, 2012).

**2.3.    Urban Stormwater Perceptions, Experiences, and Participation**

Stormwater impacts and recovery are not distributed evenly, often with the most marginalized experiencing the greatest harm and/or vulnerability. In the United States, the impoverished, unemployed, and underinsured are more likely to live in flood zones than outside, and this pattern is more prominent in inland areas than coastal zones (Qiang, 2019). These populations are at a higher risk of physical exposure to stormwater hazards. In addition to physical exposure, we also consider vulnerability to stormwater hazards, which accounts for people's ability to recover from disasters. Physical exposure and high vulnerability to stormwater hazards can lead to high stormwater risk perception in socially and economically marginalized communities.

Scholars have found that women, those with low-income, and people of color have a higher flood risk perception than men, those with high-income, and White people across multiple urban regions (Harlan et al., 2019). Higher risk perception and accounts of flood-related experiences in marginalized communities reflect their social vulnerability. Recovery from large storms can consume expendable income, and damage to transportation systems can leave many without transit to work and school—a loss of income that can be debilitating. Furthermore, gender role disparities in the private/domestic sphere lead to women bearing the brunt of flood recovery tasks. Women are often expected to care for sick and elderly family members, apply for aid from public services, and women-dominated service industries are less likely to provide job security, childcare, and uninterrupted paychecks during flood events (Enarson and Fordham, 2000; Walker and Burningham, 2011). Moreover, communities of color that are historically underserved by the government can lack trust in government-issued recovery services (Harlan et al., 2019; Pradhananga et al., 2019). This lack of trust is often a result of oppressive relationships with government officials that fail to meet the community's basic needs. Lack of equitable collaboration between institutions and communities leaves these communities isolated and vulnerable to stormwater risks.

Conventional stormwater management focused on technical solutions to flooding and water quality issues assumes that engineering approaches will result in equitable service provision (Carriquiry et al., 2020). Such a historic top-down model ignores the multiple social and environmental objectives of stormwater management and can drive a wedge between managers and the public. Some scholars have shown that water-related knowledge is positively associated with environmentally conscious behaviors suggesting that lack of knowledge can be a barrier to participation in water management (Dean et al., 2016). For instance, there is a lack of understanding about how the public's actions, like pet waste and lawn fertilizing, negatively impact water quality (Giacalone et al., 2010). Whereas others note how stormwater governance relies heavily on technical expertise, which can impede broader forms of public participation in decision-making and adoption of mitigative practices on privately owned land (Cousins, 2018). For example, stormwater agencies in Seattle, WA, Portland, OR, Philadelphia, PA, Chicago, IL, and Syracuse, NY, privileged technical expertise related to hydrological control of water and lacked formal structures for residents to participate in decision-making (Dhakal and Chevalier, 2016).

The fact that stormwater impacts and benefits are not equitably distributed calls for social and political processes to be incorporated into sustainable stormwater management programs (Hillman, 2004). There is now recognition that multiple stakeholders need to be involved in stormwater management, including residents, homeowner associations, scientists and engineers, and regulatory officials to ensure sustainable and equitable distribution of stormwater risks and benefits (Carriquiry et al., 2020). Lack of community participation in decision-making has been cited as one of the most identified barriers to building sustainable stormwater management systems (Brown and Farrelly, 2009). Community participation, especially that of the most marginalized and vulnerable individuals, is pivotal to sustainable and equitable stormwater management.

In this paper, we examined whether an individual's race, gender, and education level help predict their willingness to pay more in stormwater fees and willingness to volunteer for stream cleanups. We predicted that this work would support the environmental deprivation theory (Mohai and Bryant, 1998)—socially and racially marginalized peoples will be more exposed to stormwater hazards leading to greater concern about stormwater, and their heightened concern will lead to an intention to alleviate their conditions through participatory activities. We recognize that some environmentally conscious behaviors have higher barriers to implementation than others; therefore, we predicted that behaviors with lower barriers to action would garner more support from socially and economically marginalized peoples. For instance, environmental behaviors that require financial support will have lower buy-in from residents with lower expendable income. In contrast, activities that require hands-on participation, such as stream cleanups, directly align with other individual and community-level needs such as physical activity, environmental education, and community beautification and will garner greater support from marginalized communities.

## 3. Methods

### 3.1. Study Area and Data Collection

Three longitudinal surveys conducted in Charlotte, North Carolina, in 2014, 2016, and 2017 provide data for this study. Charlotte has a growing urban population and economy; however, in contrast to the growing prosperity, in 2015, 17% of Charlotte's residents lived below the poverty line, and Black, Native, and Latino Americans are overrepresented in this population (U.S. Census Bureau, 2015). Compared to other cities in the contiguous U.S., Charlotte has the 5th flashiest streamflow, which is indicative of the high frequency of flash flooding events (Smith and Smith, 2015). Charlotte is predicted to have a higher risk for drought and more extreme storms in future climate change scenarios (Kunkel et al., 2020). Charlotte residents who are historically marginalized by their race, gender, and class are increasingly at risk of and vulnerable to floods and stormwater pollution because they are more likely to reside in flood zones (Debbage, 2019).

The data for this paper was drawn from surveys conducted by Charlotte-Mecklenburg Storm Water Services (CMSWS), in the three years noted above (2014, 2016, and 2017), on community perception and opinion of stormwater in Mecklenburg County and the city of Charlotte. In 2014, the University of North Carolina (UNC) Charlotte's Energy and Environmental Assistance Office administered phone surveys by randomly sampling a list of purchased landline and cell phone numbers of Mecklenburg County residents. Survey administers sampled until 400 surveys were 100% complete. We do not have access to the number of attempted phone calls, and therefore we cannot calculate a response rate. The 2016 and 2017 surveys were administered by The Jackson Group, a private survey company. We accessed the data from Charlotte Stormwater Services in the summer of 2018.

**3.2.     Variables and Measures**

Details of the key measures used to operationalize each variable in our analysis are discussed below. The complete list of survey measures can be found in Appendix 1.

*Outcome Variables*

We used three measures as outcomes: *willingness to volunteer*, *willingness to pay*, and *concern for flooding*. Willingness to volunteer is a single item related to willingness to clean up a local stream. Willingness to pay is a summed scale of two item measures ($\alpha_{2014}$= 0.69; $\alpha_{2016}$= 0.83; $\alpha_{2017}$= 0.85; Cronbach's α is a measure of internal consistency). Each item relates to willingness to pay more in stormwater fees to improve flooding or water quality. Willingness to volunteer was only measured in 2014, and willingness to pay was measured in 2014, 2016, and 2017. *Concern for flooding* is a summed scale of two item measures that represent respondents' concern with local flooding of buildings and roads ($\alpha_{2014}$= 0.77; $\alpha_{2016}$= 0.77; $\alpha_{2017}$= 0.69). These variables were originally measured on a Likert-scale (1= don't know, 2= strongly disagree, 3= disagree somewhat, 4= agree somewhat, 5=agree strongly). We recoded all variables to a 4-point scale (1= strongly disagree, 2= disagree somewhat, 3= agree somewhat, 4=agree strongly) prior to analysis. "Don't know" responses were not included in our analysis.

*Predictor Variables*

The predictor variables of interest in the study are *race/ethnicity*, *gender*, and *education level*. Race/ethnicity was measured as a nominal variable— including non-Latino White (reference level), Latino of any race, non-Latino Black/African American, non-Latino Asian American and Pacific Islander, non-Latino multi-racial, and other race/ethnicity. Gender was measured as a dichotomous variable—female (reference level) and male. Education level was measured as an ordinal variable that varies from "less than high school" to "graduate study."

Control variables are *exposure to stormwater ads, knowledge of stormwater, age, homeownership, residence in a flood zone,* and *time*. Exposure to stormwater ads is a summed scale of five item measures related to whether respondents have seen or heard recent Charlotte stormwater advertisements ($\alpha_{2014}$= 0.61; $\alpha_{2016}$= 0.83; $\alpha_{2017}$= 0.80). This scale measures exposure to informal awareness-raising campaigns that often have short and digestible messages to the public about stormwater. CMSWS runs educational advertisements on stormwater and flood awareness, such as the "Turn Around, Don't Drown" campaign. Stormwater advertisement campaigns were not the same content-wise across all years; therefore, exposure to advertisements represents a count of advertisements that a respondent has seen in a given year. Knowledge of stormwater is a measure of a respondent's technical knowledge of stormwater treatment. We developed a summed scale of two items representing awareness that stormwater directly drains to local streams without treatment facilities ($\alpha_{2016}= 0.64$; $\alpha_{2017}= 0.61$). Knowledge of stormwater was only considered in our analysis for 2016 and 2017 due to inconsistent item measurements, poorly worded survey items, and lack of reliability in 2014.

We replaced missing values in the independent variables if the variables had less than 10% of values missing. We replaced missing values for education level, exposure to stormwater ads, age, residence in a flood zone, homeownership, and knowledge of stormwater. We performed multiple imputations using chained equations to replace "missing at random" (MAR) values of the independent variables (Graham, 2009).

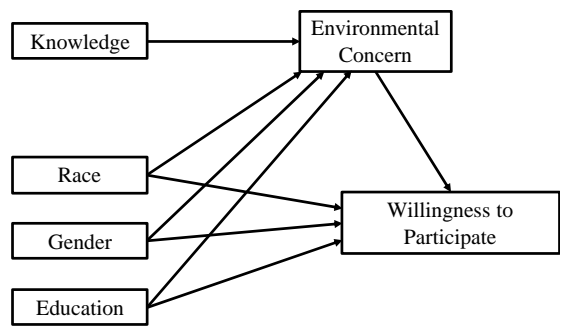

**Figure 1.** Conceptual model of willingness to participate based on a simple attitude-behavior model. An individual's knowledge of stormwater informs their concern about stormwater, and concern about stormwater influences their willingness to participate. Using this framework, we test for direct and indirect influences of an individual's race, gender, and education level on their willingness to participate.

### 3.3. Modeling and Analytic Strategy

Data were analyzed using R Programming (R Core Team, 2013) and PROCESS in SPSS Statistics 26 (Hayes, 2017). First, we examined whether socially marginalized groups reported higher exposure to flood zones using a test of equal proportions. Two additional objectives were assessed using mediation analysis: 1) the direct and indirect effects of race, gender, and education level on an individual's willingness to participate, and 2) whether the type of participation influences the association between participation and race, gender, and education.

We used a simple attitude-behavior model to frame our investigation. Attitudes influence behavioral intentions (Ajzen, 1991), and knowledge as well as correlates of knowledge modify attitudes rather than directly influence behavioral intentions (Kollmuss and Agyeman, 2002). In this framework, knowledge (knowledge of stormwater and exposure to educational advertisements) informs attitudes (concern for flooding), and attitudes predict willingness to participate (Figure 1). Then, we used a more exploratory approach, rooted in empirical evidence of the predictive pathways, to test for direct and indirect relationships between race, gender, and education level and willingness to participate.

Prior research has shown that an individual's race, gender, and education level influence *both* environmental attitudes and intentions to participate in environmentally conscious behaviors, and path models allowed us to investigate both direct and indirect pathways. We tested for two possibilities: (1) race, gender, and education level have an indirect effect on willingness to participate, which implies that their effect is mediated by differential concerns for flooding; or (2) the effect is predominantly direct, which implies that the effect is largely unrelated to one's concern for flooding. Empirical studies point towards the first explanation (Botetzagias et al., 2015); however, a significant direct effect can also imply that we did not capture the specific attitude (an unmeasured mediator) that mediates the relationship between individuals' race, gender, and education and their willingness to participate.

We performed mediation analysis via OLS (Ordinary Least Squares) path analysis— a causal model in which a series of multiple regressions are estimated to examine the effect of a set of variables on a specified outcome via multiple mechanisms. The general equations of an OLS path analysis are as follows (Hayes, 2017):

$$M = i_M + a_1X_1 + a_2X_2 + \ldots + a_kX_k + e_M \tag{1}$$

$$Y = i_Y + c_1X_1 + c_2X_2 + \ldots + c_kX_k + bM + e_Y \tag{2}$$

where $M$ is the mediating effect, $i$ is the constant, $X_i$ are predictor variables, $e$ is error, and $Y$ is the outcome variable. Equations 1 and 2 estimate the direct effect coefficients that predict $M$ and $Y$ outcome variables. The indirect effect of $X$ on $Y$ through $M$ is the product of two effects: $a_i b$. We investigated two paths: one where race, gender, and education directly influence respondents' willingness to participate, and another where race, gender, and education indirectly influence respondents' willingness to participate through a mediator variable, concern for flooding. While we included all independent variables and controls in the model, we only calculated the indirect effect of race, gender, and education level as these are the predictor variables of interest. We estimated bootstrapped confidence intervals for the indirect effect based on 5,000 bootstrapped samples of the indirect effects. We performed two mediation analyses with our outcome variables of interest: willingness to pay and willingness to volunteer. Willingness to volunteer was only assessed in the 2014 survey, and willingness to pay utilizes a pooled dataset including 2014, 2016, and 2017 surveys. We ran an additional path analysis with a pooled dataset, including 2016 and 2017 samples, because this dataset has a reliable measure of stormwater knowledge.

To ensure that our model choice was a good fit, we checked the assumptions of multiple regression (Appendix 2). We also compared the results of the multiple regression models to proportional odds models to ensure that our assumption to treat ordinal response variables as continuous would not significantly influence our results (Appendix 3).

**4.     Results**

**4.1.     Sample Characteristics**

Respondents in the 2014 sample resembled the demographics of Mecklenburg County in terms of race (47% White, 13% Latino of any race, 31% Black, 6% Asian, 2% Multi-racial), gender (52% Female), and education level (average educational attainment is "some college or associate's degree") (U.S. Census Bureau, 2019). In 2016 and 2017, there was a higher proportion of White respondents (64%) compared to 2014 (58%), and Latino, Black, and Asian American respondents were underrepresented (Table 1). The 'Other' racial category includes individuals who refused to respond to the race and ethnicity questions and individuals belonging to racial/ethnic groups with cumulatively fewer than 20 representatives across the three surveys. Gender and education level of the respondents were representative of the population in all surveys. The mean age of survey respondents increased over time from 35-44 to 45-54 years old. Additionally, in 2014, 70% of respondents were homeowners, which is higher than the city average of 56% (U.S. Census Bureau, 2019).

### 4.2. Flood Zone Residence

A test of equal proportions revealed that racially marginalized respondents were significantly overrepresented in flood zones (Figure 2). Non-White respondents represented roughly half of the residents that claimed to live in a flood zone, while they were 30% of the population outside of flood zones. There were no significant differences in the proportion of respondents that resided in flood zones compared to that outside of flood zones by gender and education level.

### 4.3. Dependent Variable 1: Willingness to Volunteer

The mediation analysis indicated that racially marginalized respondents, on average, were more concerned for flooding and more willing to volunteer than White respondents; however, respondents' concern for flooding was not significantly associated with their willingness to volunteer. Latino, Black, and multi-racial respondents were more concerned for flooding near their homes and businesses compared to White respondents (reference level) when all other variables are held constant ($a = 1.34, S.E. = 0.39, p < 0.01$; $a = 0.84, S.E. = 0.23, p < 0.01$; $a = 0.86, S.E. = 0.47, p = 0.07$, respectively; Table 2). Latino respondents, on average, were more willing to volunteer for stream cleanups than White respondents ($c = 0.84, S.E. = 0.22, p < 0.01$; Table 2). However, respondents' concern for flooding was not significantly associated with their willingness to volunteer ($b = 0.04, S.E. = 0.03, p = 0.15$; Table 2), resulting in a predominantly direct effect of race on willingness to volunteer. This indicates that the heightened willingness to volunteer in Latino respondents was not associated with their concern for flooding. We did not find a consistent response across all racial groups. Asian American and Pacific Islander respondents were, on average, just as concerned for flooding as

**Table 1.** Sample characteristics in 2014, 2016, and 2017 surveys.

|  | 2014 | | | | 2016 | | | | 2017 | | | |
| --- | --- | --- | --- | --- | --- | --- | --- | --- | --- | --- | --- | --- |
|  | Range | Mean | SD | N | Range | Mean | SD | N | Range | Mean | SD | N |
| *Outcome Variables* | | | | | | | | | | | | |
| Willingness to Pay | 1-7 | 4.86 | 1.87 | 402 | 1-7 | 3.69 | 1.83 | 393 | 1-7 | 3.57 | 1.82 | 363 |
| Willingness to Volunteer | 1-4 | 3.02 | 1.06 | 403 | | | | | | | | |
| Concern for flooding | 1-7 | 4.28 | 1.95 | 397 | 1-7 | 4.90 | 1.49 | 409 | 1-7 | 4.80 | 1.38 | 394 |
| *Predictor Variables* | | | | | | | | | | | | |
| Race | | | | | | | | | | | | |
| White | 0-1 | 0.58 | | 233 | 0-1 | 0.64 | | 264 | 0-1 | 0.64 | | 255 |
| Latino | 0-1 | 0.07 | | 30 | 0-1 | 0.07 | | 29 | 0-1 | 0.03 | | 12 |
| Black/African American | 0-1 | 0.24 | | 98 | 0-1 | 0.18 | | 76 | 0-1 | 0.11 | | 44 |
| Asian or Pacific Islander | 0-1 | 0.05 | | 20 | 0-1 | 0.04 | | 15 | 0-1 | 0.02 | | 9 |
| Multi-racial | 0-1 | 0.04 | | 17 | 0-1 | 0.02 | | 7 | 0-1 | 0.03 | | 12 |
| Other | 0-1 | 0.01 | | 5 | 0-1 | 0.05 | | 22 | 0-1 | 0.17 | | 68 |
| Gender (Female = 0) | 0-1 | 0.47 | | 402 | 0-1 | 0.52 | | 413 | 0-1 | 0.52 | | 359 |
| Education | 1-5 | 3.61 | 1.11 | 403 | 1-5 | 3.67 | 1.08 | 413 | 1-5 | 3.83 | 1.05 | 359 |
| *Control Variables* | | | | | | | | | | | | |
| Exposure to ads | 1-6 | 2.77 | 1.43 | 403 | 1-6 | 2.37 | 1.71 | 412 | 1-6 | 2.05 | 1.51 | 385 |
| Knowledge of stormwater | | | | | 1-3 | 2.46 | 0.75 | 411 | 1-3 | 2.47 | 0.75 | 395 |
| Age | 1-6 | 3.96 | 1.55 | 402 | 1-6 | 4.37 | 1.47 | 413 | 1-6 | 4.63 | 1.39 | 358 |
| Residence in Flood Zone | 0-1 | 0.06 | | 403 | 0-1 | 0.04 | | 413 | 0-1 | 0.05 | | 390 |
| Homeownership (Rent =0) | 0-1 | 0.70 | | 403 | | | | | | | | |

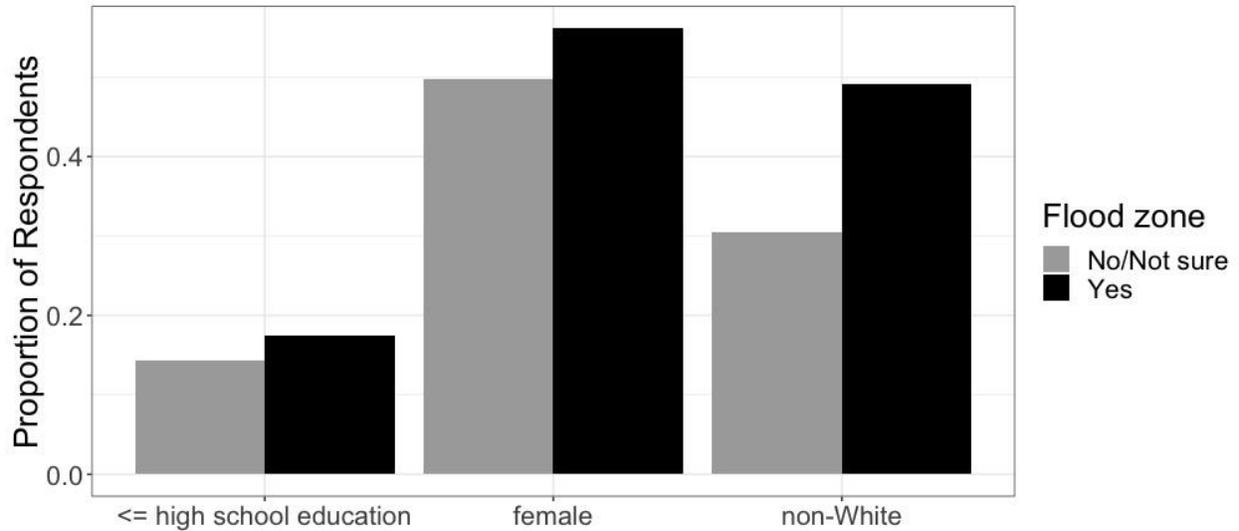

**Figure 2**. Proportion of respondents who reported residence in a flood zone by sociodemographic variables. We performed a test of equal proportions to assess significant differences. Non-White respondents are significantly overrepresented in flood zone residences (49% vs. 30%, $p < 0.01$). There were no significant differences between the proportions of women within and outside of flood zones (56% vs. 48%, $p = 0.42$) and the proportions of respondents with less than or equal to high school education level within and outside of flood zones (18% vs. 14%, $p = 0.63$).

**Table 2**. OLS regression coefficients. Following equations (1) and (2), concern for flooding is the mediating variable ($M$), and willingness to volunteer (*Volunteer*) is the outcome variable ($Y$). Standard errors of the direct effect are presented in parentheses. Indirect effect coefficients ($a_i b$) are presented with the bootstrapped 95% confidence interval in brackets.

| Independent Variable (IV) | Direct Effects | | | | Indirect Effects |
|---|---|---|---|---|---|
| | IV → Concern | | IV → Volunteer | | IV → Concern → Volunteer |
| | $a$ | $p$ | $c$ | $p$ | $a_i b$ |
| Race (White = 0) | | | | | |
| Latino | 1.34 (0.39) | <0.01*** | 0.84 (0.22) | <0.01*** | 0.05 [-0.03, 0.15] |
| Black/African American | 0.84 (0.23) | <0.01*** | 0.03 (0.13) | 0.83 | 0.03 [-0.02, 0.09] |
| Asian or Pacific Islander | 0.35 (0.44) | 0.42 | 0.01 (0.24) | 0.99 | 0.01 [-0.02, 0.06] |
| Multi-racial | 0.86 (0.47) | 0.07* | -0.01 (0.26) | 0.96 | 0.03 [-0.02, 0.11] |
| Other | -0.47 (0.83) | 0.58 | 0.03 (0.46) | 0.95 | -0.02 [-0.12, 0.07] |
| Gender (Female = 0) | -0.10 (0.19) | 0.60 | 0.08 (0.11) | 0.45 | -0.004 [-0.03, 0.01] |
| Education | -0.17 (0.09) | 0.06* | 0.07 (0.05) | 0.18 | -0.007 [-0.02, 0.004] |
| Age | 0.15 (0.07) | 0.03** | -0.11 (.04) | <0.01*** | |
| Flood zone | 1.11 (0.40) | <0.01*** | 0.55 (0.22) | 0.01*** | |
| Homeownership (Rent = 0) | -0.38 (0.23) | 0.09* | 0.08 (0.12) | 0.51 | |
| Exposure to ads | 0.25 (0.07) | <0.01*** | -- | -- | |
| Concern for flooding ($b$) | -- | | 0.04 (0.03) | 0.15 | |
| $R^2$ | 0.14 | | 0.10 | | |
| MSE | 3.36 | | 1.03 | | |
| F Statistic | 5.89*** (df = 11) | | 3.68*** (df = 11) | | |
| N | 396 | | 396 | | |

*Note:* *p<0.1; **p<0.05; ***p<0.01
**Bold** indicates that the confidence interval for the indirect estimate does not contain zero
Letters *a, b*, and *c* indicate coefficients displayed in Equations (1) and (2)

White respondents ($a = 0.35, S.E. = 0.44, p = 0.42$; Table 2). Those with lower education, on average, were more concerned for flooding ($a = -0.17, S.E. = 0.09, p = 0.06$). Gender was not significantly associated with respondents' concern for flooding or willingness to volunteer ($a = -0.10, S.E. = 0.19, p = 0.60; c = 0.08, S.E. = 0.11, p = 0.45$, respectively; Table 2).

Our results illustrate that race and education level were associated with respondents' concern for flooding and willingness to volunteer after controlling for other factors. Among the controls, concern for flooding was significantly higher among older respondents, those living within a flood zone, renters, and respondents who have seen more stormwater advertisements (Table 2). In Charlotte, stormwater fees are decided by the amount of impervious land on one's property; therefore, it is likely that homeowners pay more in stormwater fees than renters. Also, homeowners receive stormwater bills directly, while for some renters, water and sewage fees are included in the rental payment. Willingness to volunteer was significantly higher among respondents living within a flood zone and younger respondents (Table 2).

### 4.4. Dependent Variable 2: Willingness to Pay

We found that racially marginalized respondents were more concerned for flooding, and, in contrast to willingness to volunteer, respondents' concern for flooding was positively associated with their willingness to pay increased stormwater fees (Table 3). On average, Latino, Black, and multi-racial respondents were more concerned about flooding compared to White respondents when all other variables are held constant ($a = 1.18, S.E. = 0.22, p < 0.01; a = 0.51, S.E. = 0.13, p < 0.01; a = 0.49, S.E. = 0.27, p = 0.07$, respectively; Table 3). In turn, respondents who were more concerned about flooding expressed a greater willingness to pay ($b = 0.24, S.E. = 0.03, < 0.01$). Results also indicate that the effect of race on willingness to pay was fully mediated by concern for flooding. Full mediation occurs when an indirect effect is present without a significant direct effect (Zhao et al., 2010), and in this case, indicates that Latino and Black respondents' heightened concern for flooding was significantly and positively associated with their willingness to pay. We also found that women (reference level) and those with lower education, on average, were more concerned about flooding ($a = -0.22, S.E. = 0.10, p = 0.02$; $a = -0.15, S.E. = 0.05, p < 0.01$, respectively; Table 3), and, in turn, their concern for flooding was positively associated with their willingness to pay. The effect of gender on willingness to pay was fully mediated by concern for flooding, as indicated by the presence of an indirect effect without a significant direct effect.

We observed competitive mediation in reference to education level: direct ($c = 0.08, S.E. = 0.05, p = 0.11$; Table 3) and indirect ($ab = -0.04, [upper\ limit, lower\ limit] = [-0.06, -0.01]$; Table 3) effects exist but in opposite directions (Zhao et al., 2010). This result suggests multiple mechanisms by which education influences a respondents' willingness to pay. The indirect effect suggests that respondents with lower education were more concerned about flooding, and their concern positively influenced their willingness to pay. After controlling for the indirect effect of concern for flooding, there remains an effect of education on willingness to pay. This effect works in the opposite direction: those with higher education were more willing to pay, but their concern for flooding did not drive their willingness to pay. Of the controls, older respondents, those living within flood zones, and respondents with greater exposure to stormwater advertisements were more concerned about flooding. Willingness to pay was significantly higher among younger respondents (Table 3). Interestingly, we found that concern for flooding increased over time ($a = 0.25, S.E. = 0.04, p < 0.01$), but willingness to pay decreased over time ($c = -0.43, S.E. = 0.05, p < 0.01$).

**Table 3**. OLS regression coefficients. Following equations (1) and (2), concern for flooding is the mediating variable ($M$), and willingness to pay ($WTP$) is the outcome variable ($Y$). Standard errors of the direct effect are presented in parentheses. Indirect effect coefficients ($a_i b$) are presented with the bootstrapped 95% confidence interval in brackets.

| Independent Variable (IV) | Direct Effects | | | | Indirect Effects |
| --- | --- | --- | --- | --- | --- |
| | IV → Concern | | IV → WTP | | IV → Concern → WTP |
| | $a$ | p | $c$ | p | $a_i b$ |
| Race (White = 0) | | | | | |
| Latino | 1.18 (0.22) | <0.01*** | 0.02 (0.25) | 0.95 | **0.28 [0.15, 0.43]** |
| Black/African American | 0.51 (0.13) | <0.01*** | 0.11 (0.15) | 0.45 | **0.12 [0.05, 0.20]** |
| Asian or Pacific Islander | 0.27 (0.25) | 0.27 | 0.01 (0.28) | 0.98 | 0.07 [-0.04, 0.18] |
| Multi-racial | 0.49 (0.27) | 0.07* | -0.29 (0.31) | 0.35 | 0.12 [-0.03, 0.27] |
| Other | -0.05 (0.24) | 0.83 | -0.56 (0.27) | 0.04** | -0.01 [-0.13, 0.10] |
| Gender (Female = 0) | -0.22 (0.10) | 0.02** | -0.06 (0.11) | 0.60 | **-0.05 [-0.10, -0.007]** |
| Education | -0.15 (0.05) | <0.01*** | 0.08 (0.05) | 0.11 | **-0.04 [-0.06, -0.01]** |
| Age | 0.10 (0.03) | <0.01*** | -0.22 (0.04) | <0.01*** | |
| Flood zone | 0.69 (0.22) | <0.01*** | 0.05 (0.25) | 0.83 | |
| Time | 0.25 (0.04) | <0.01*** | -0.43 (0.05) | <0.01*** | |
| Exposure to ads | 0.09 (0.03) | <0.01*** | -- | | |
| Concern for flooding ($b$) | -- | | 0.24 (0.03) | <0.01*** | |
| $R^2$ | 0.10 | | 0.16 | | |
| MSE | 2.50 | | 3.18 | | |
| F Statistic | 11.64*** (df = 11) | | 18.42*** (df = 11) | | |
| N | 1115 | | 1115 | | |

*Note:* *p<0.1; **p<0.05; ***p<0.01
**Bold** indicates that the confidence interval for the indirect estimate does not contain zero
Letters *a, b*, and *c* indicate coefficients displayed in Equations (1) and (2)

We also conducted an OLS path analysis with the pooled samples from 2016 and 2017, excluding data from 2014. The purpose of this analysis was to test the influence of knowledge about stormwater on respondents' concern for flooding and willingness to pay. The results of this analysis were similar to the findings reported in Table 3 and additionally illustrated that knowledge about stormwater was not significantly associated with respondents' concern for flooding ($a = 0.08, S.E. = 0.07, p = 0.28$; Appendix 4).

## 5. Discussion

In this paper, we examine if and how an individual's gender, race, and education level help predict their concern for flooding and willingness to participate in stormwater management. Consistent with previous literature (Lazri and Konisky, 2019; Macias, 2016a), we found that racially marginalized individuals, women, and those with a lower education level reported higher concern for local flooding compared to White, male, and higher educated respondents. Moreover, a heightened concern for flooding was an essential pathway through which socially and economically marginalized individuals developed their increased willingness to participate. The racial disparity in concern for flooding was greatest between White and Latino participants as Latinos were, on average, 1.2 units higher in their concern for flooding on a scale from 1-7 (Table 3). The disparity in concern for flooding and willingness to participate remained even after considering the effects of flood zone residence and stormwater knowledge and awareness.

We illustrate that racial disparities exist in flood zone residence, with non-White respondents being overrepresented in flood zones (Figure 2). Likewise, a recent study showed that Black, Latino, and impoverished communities in Mecklenburg County are more likely to reside in flood zones than White and wealthier populations (Debbage, 2019). Our work takes this one step further in that racial disparities in concern for flooding and willingness to participate remain even after considering flood zone residence (Tables 2 and 3). Our results suggest that it is not only physical risk exposure but also differential vulnerability to stormwater hazards that drives risk perceptions of flood-related hazards (Hale et al., 2018). Like physical exposure, vulnerability to hazards is also driven by structural forms of racism, classism, and sexism that create and sustain debilitating patterns of unequal wealth distribution, access to loans, and access to transportation (Masozera et al., 2007). As Jones and Rainey (2006) pointed out, a perceived lack of social and financial support to address environmental hazards can trigger concerns about one's capacity to adapt. Importantly, the link we show between race, concern for flooding, and willingness to participate suggests that heightened concerns from racially marginalized individuals can lead to intentions to improve stormwater management.

Our results also highlight the complexity in willingness to pay measures, which tap into economic opportunity. We found that less-educated individuals had higher concerns for flooding and were, in turn, more willing to pay to improve flooding. However, education level still explained a portion of willingness to pay that was independent of concern for flooding: people with higher education were more willing to pay to improve flooding, which aligns with previous empirical studies (Chui and Ngai, 2016; Dietz et al., 2007; Macias, 2016a; Newburn and Alberini, 2016). Identifying such relationships highlights the complexity of willingness to pay measures, especially when associated with education level. High exposure to flooding hazards can drive low-income individuals' concern about stormwater and thus increase their willingness to participate in improving their conditions; however, a lack of expendable income or access to pertinent resources may deter willingness to participate in activities involving payment for services. Notably, this association was less evident with regard to willingness to volunteer (Table 2), implying that barriers that exist for willingness to pay may not substantially deter willingness to volunteer. These results highlight the need for multiple measures of environmentally conscious participation that capture the complexities of behaviors and barriers to performing them.

Lack of knowledge about stormwater has been cited as a key barrier to participation; however, we found an inconsistent effect of knowledge about stormwater on concern for flooding and willingness to participate. While technical knowledge about stormwater was not associated with concern for flooding, exposure to educational advertisements was positively associated with concern for flooding. Our results suggest that technical expertise of watershed processes does not influence concern about stormwater; rather, as Mobley (2016) pointed out, informal education—driven by educational campaigns and everyday experiences with and observations of flooding and water quality— likely drives concern about and participation in stormwater management. This finding has implications for understanding the importance of environmental experiences, rather than formal knowledge, on participation in water management. For example, pervasive segregation of U.S. cities, driven by legacies of discriminatory housing segregation policies, can dictate differential environmental experiences by socioeconomic status. These differences in experiences possibly facilitate different understandings of flooding and water quality impacts and thus different concerns about and participation in stormwater management.

Future research will be needed to further explore the implications of the current study. First, future work could explore how sociodemographic variables can be included in classic models of environmental behavior, such as the Theory of Planned Behavior (Ajzen, 1991). Given the growing evidence that race, gender, and class significantly influence concern and behavior, more researchers should aim to sample and gather data in representative ways along these lines.

Second, our measurement of willingness to participate is limited by our ability to only consider willingness to pay and volunteer. We observe differences in how these activities are related to race, gender, education, and age, which calls for researchers to expand measurements of participation and willingness to participate to include other activities such as reporting infrastructure failures to local agencies and joining advocacy organizations. This research would also benefit by measuring actual behavior rather than the intent to participate. By measuring actual levels of participation, we can develop a more comprehensive understanding of barriers to action. Third, given our results, we should expand our conceptualization of "knowledge" in survey instruments. As shown in this study, technical expertise of watershed processes does not influence concern about flooding, but targeted campaigns like CMSWS "Turn Around, Don't Drown" and people's experiences with local flooding are more influential.

There are some notable shortcomings of the surveys used in our analysis. First, each year that the survey was conducted, administrators sampled until 400 surveys were complete. There was a lack of attention towards survey response rates, which signals low response rates and likely biased the survey participants towards those who are more opinionated about stormwater. Second, in the 2014 survey, respondents' gender was recorded without explicitly asking their gender identity. Survey administrators determined gender by the voice of the respondents. This practice not only introduced biases into the gender measurement but also stripped participants of agency to define their gender. This practice was not repeated in 2016 and 2017. Third, while our sample yielded enough cases to be a valid representation of racial groups, Black, Latino, and Asian American respondents were underrepresented compared to their representation in Mecklenburg county. Future surveys should address this bias by conducting stratified random sampling to ensure multiple population characteristics are represented in the sample. Fourth, we recommend that future surveys measure respondents' experiences with flooding because it can be an important predictor of concerns for flooding (*see* Hale et al., 2018). Urban flooding is spatially heterogeneous due to stormwater infrastructure and impervious surfaces and can deviate from riverine floodplains considerably. Given that flood zones are not well known, flooding experiences could be a more accurate measurement of flood exposure than residence within a 100-year floodplain. Despite the shortcomings mentioned above that often come with secondary data, we find that these surveys provide unique and timely information about willingness to participate in stormwater management based on an individual's race, gender, and education level.

## 6. Concluding Thoughts

We evaluated the relationship between social and economic marginalization and stormwater management. We found consistent and significant associations between race, gender, and education level and individuals' concern about and willingness to participate in stormwater management across three longitudinal surveys. More underserved groups were more willing to participate in stormwater remediation, and willingness to participate was associated with their heightened concern for flooding. These analyses have considerable implications for how we theorize the interplay between inequitable distributions of environmental conditions and actions to remediate those conditions. Even in highly technical spaces where public participation is unconventional, we saw that socially and racially marginalized individuals were driven by their concerns for flooding to be more involved in shaping their future relationship with stormwater. These patterns should alert policymakers to recognize this heightened concern and facilitate ways in which these communities can articulate their experiences and be involved in stormwater decision-making and planning.

**Acknowledgments**

We would like to acknowledge Charlotte-Mecklenburg Storm Water Services for providing us with the survey data, meeting with us to discuss the study's preliminary results, and their generous support of this research.

**Funding**

We acknowledge the travel grant provided by the Department of Agricultural and Biological Engineering at Purdue University.

APPENDIX 1— SURVEY VARIABLES AND MEASURES

| Variables | Questions |
|---|---|
| Willingness to Pay[a]<br>Summed Scale 1 ($\alpha_{2014} = 0.69$) | I would be willing to pay more in storm water fees if it would be used to clean up polluted creeks and streams in Charlotte-Mecklenburg.<br>I would be willing to pay more in storm water fees if it would reduce flooding in Charlotte-Mecklenburg. |
| Willingness to Pay[a]<br>Summed Scale 2 ($\alpha_{2016} = 0.83$; $\alpha_{2017} = 0.85$) | I would be willing to pay more in stormwater fees if it would be used to clean up polluted creeks, streams, and lakes in Charlotte and Mecklenburg County<br>I would be willing to pay more in stormwater fees if it would be used to reduce flooding in Charlotte and Mecklenburg County. |
| Willingness to Volunteer[a] | I would volunteer to work with a group of volunteers, twice a year, to help clean up polluted creeks and streams in or near my neighborhood. |
| Concern for Flooding[a]<br>Summed Scale 1 ($\alpha_{2014} = 0.77$) | During times of heavy rain, I am concerned that the creeks in Charlotte-Mecklenburg will flood roads.<br>During times of heavy rain, I am concerned that the creeks in Charlotte-Mecklenburg will flood buildings |
| Concern for Flooding[a]<br>Summed Scale 2 ($\alpha_{2016} = 0.77$; $\alpha_{2017} = 0.69$) | During times of heavy rain I am concerned about stormwater flooding on roads.<br>During times of heavy rain I am concerned about flooding in homes and buildings. |
| Exposure to stormwater advertisements[b]<br>Summed Scale 1 ($\alpha_{2014} = 0.61$) | Have you heard or seen any information about storm water in the past year?<br>Advertisements have been run on multiple mediums letting people know it's so easy to volunteer. Do you recall seeing or hearing any of these ads in the past year?<br>Advertisements have been run on multiple mediums encouraging citizens to Turn around Don't drown with the tag line your car is not a boat. Do you recall seeing or hearing any of these ads in the past year? |
| Exposure to stormwater advertisements[b]<br>Summed Scale 2 ($\alpha_{2016} = 0.83; \alpha_{2017} = 0.80$) | Local emergency responders and flood experts have started an educational effort called 'Build an Ark' about flood awareness, responsibility and knowledge. Have you heard or seen anything about the Build an Ark campaign?<br>Advertisements have been run on multiple mediums encouraging citizens to be a water watcher and report storm water or creek pollution. Do you recall seeing or hearing any of these ads in the past year?<br>Have you heard or seen any information from Charlotte-Mecklenburg Storm Water Services about flood zone maps and flood insurance in the past 12 months?<br>Have you heard or seen any information from Charlotte-Mecklenburg Storm Water Services about volunteer opportunities to reduce pollution in and around our creeks, streams, and lakes in the past 12 months?<br>Have you heard or seen any information from Charlotte-Mecklenburg Storm Water Services about reporting pollution in stormwater, creeks, streams, and lakes, such as the "Water Watcher" program, in the past 12 months?<br>Have you heard or seen any information from Charlotte-Mecklenburg Storm Water Services about stormwater pollution and how it flows to creeks, streams, and lakes in the past 12 months?<br>Have you heard or seen any information from any other entity about stormwater pollution, flooding, volunteer opportunities, etc.? |
| Knowledge of Stormwater[a]<br>Summed Scale 1 ($\alpha_{2014} = 0.12$) | Water that runs into storm drains is treated and cleaned before being released into creeks, lakes and ponds.<br>People in Charlotte-Mecklenburg can experience severe flooding even if their property is not in a flood zone.<br>Water that runs into storm drains flows directly to local creeks and lakes. |
| Knowledge of Stormwater[c]<br>Summed Scale 2 ($\alpha_{2016} = 0.64; \alpha_{2017} = 0.61$) | Stormwater that goes into storm drains is cleaned by a treatment facility before it goes into creeks, streams, and lakes.<br>Water that flows into storm drains goes directly to local creeks, streams, and lakes. |
| Residence in a Flood Zone[b] | Is the property where you currently live in a flood zone? |

| Home Ownership[d] | Do you own or rent your home? |
|---|---|

*Notes on coding:*

[a] 1 = Don't know, 2 = Disagree strongly, 3 = Disagree somewhat, 4 = Agree somewhat, 5 = Agree strongly

[b] 1 = Yes, 2 = No, 3 = Don't know

[c] 1 = False, 2 = True

[d] 1 = Own, 2 = Rent, 3 = Other, 4 = Refused to answer

APPENDIX 2 – ASSUMPTIONS OF MULTIPLE REGRESSION

We checked multiple regression assumptions for each model displayed in the main text (Tables 2-4). We checked the assumptions of the direct effect model with participation (willingness to volunteer, willingness to pay, and overall willingness) as the outcome variable. We checked for multicollinearity using the generalized variance inflation factor (GVIF), normality using a histogram of the regression residuals, and homoscedasticity by plotted residual vs. fitted values of the regression model.

**Table 1.** The generalized variance inflation factor for predictor variables in the regression displayed in Table 2 of the main text. *Willingness to Volunteer* is the outcome variable.

| Variable | GVIF | GVIF$^{(1/2Df)}$ |
|---|---|---|
| Race | 1.46 | 1.04 |
| Education | 1.24 | 1.12 |
| Gender | 1.05 | 1.03 |
| Exposure to Ads | 1.19 | 1.09 |
| Concern for Flooding | 1.17 | 1.08 |
| Home Ownership | 1.26 | 1.12 |
| Age | 1.29 | 1.13 |
| Exposure to flood zone | 1.07 | 1.04 |

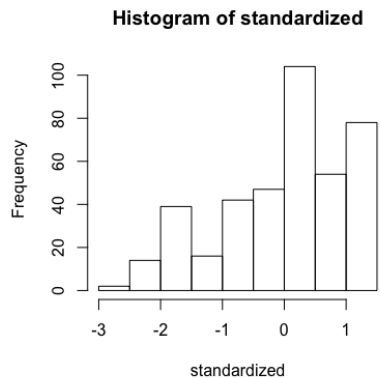

**Figure 1.** Histogram of the standardized residuals of the regression displayed in Table 2 of the main text. *Willingness to Volunteer* is the outcome variable.

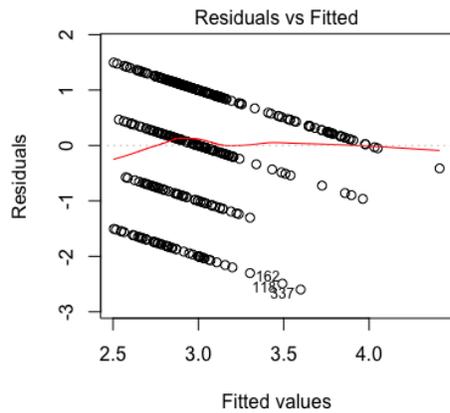

**Figure 2.** Residuals vs. fitted values of the regression displayed in Table 2 of the main text. *Willingness to Volunteer* is the outcome variable.

**Table 2.** The generalized variance inflation factor for predictor variables in the regression displayed in Table 3 of the main text. *Willingness to Pay* is the outcome variable.

| Variable | GVIF | GVIF$^{(1/2Df)}$ |
|---|---|---|
| Race | 1.32 | 1.03 |
| Education | 1.13 | 1.06 |
| Gender | 1.03 | 1.02 |
| Exposure to Ads | 1.09 | 1.04 |
| Concern for Flooding | 1.12 | 1.06 |
| Time | 1.17 | 1.04 |
| Age | 1.17 | 1.08 |
| Exposure to flood zone | 1.03 | 1.01 |

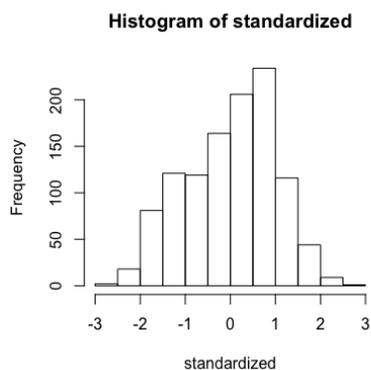

**Figure 3.** Histogram of the standardized residuals of the regression displayed in Table 3 of the main text. *Willingness to Pay* is the outcome variable.

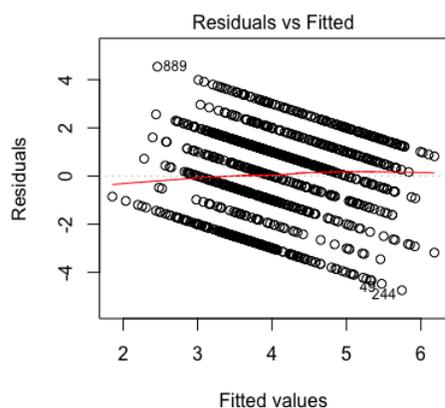
**Figure 4.** Residuals vs. fitted values of the regression displayed in Table 3 of the main text. *Willingness to Pay* is the outcome variable.

**Table 3.** The generalized variance inflation factor for predictor variables in the regression displayed in Appendix 4. *Willingness to Pay* is the outcome variable.

| Variable | GVIF | GVIF$^{(1/2Df)}$ |
|---|---|---|
| Race | 1.32 | 1.03 |
| Education | 1.14 | 1.07 |
| Gender | 1.07 | 1.03 |
| Exposure to Ads | 1.05 | 1.03 |
| Concern for Flooding | 1.08 | 1.04 |
| Time | 1.03 | 1.02 |
| Exposure to flood zone | 1.02 | 1.01 |
| Age | 1.15 | 1.07 |
| Knowledge | 1.07 | 1.03 |

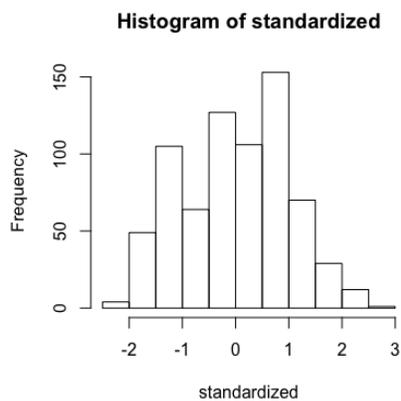

**Figure 5.** Histogram of the standardized residuals of the regression displayed in Appendix 4. *Willingness to Pay* is the outcome variable.

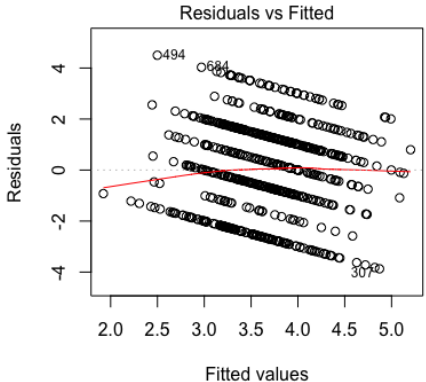

**Figure 6.** Residuals vs. fitted values of the regression displayed in Appendix 4. *Willingness to Pay* is the outcome variable.

APPENDIX 3 – PROPORTIONAL ODDS MODEL

**Table 1.** We performed several proportional odds models to test whether our assumption of continuous response variables significantly impacted the model results. This model is comparable to Table 2 in the main body of the text. The outcome variable is willingness to volunteer. The significant variables and the direction of the effect is the same in both the proportional odds model shown below and the multiple regression shown in Table 2 of the main text.

| Independent Variable | Odds Ratio |
|---|---|
| Race | |
| Latino | 10.87*** |
| Black/African American | 1.14 |
| Asian or Pacific Islander | 0.96 |
| Multi-racial | 0.85 |
| Other | 1.58 |
| Gender (Female = 0) | 1.19 |
| Education | 1.14 |
| Age | 0.84** |
| Flood zone | 3.36*** |
| Home Ownership (Rent = 0) | 1.09 |
| Exposure to ads | 0.99 |
| Concern for flooding | 1.09 (p = 0.12) |
| Observations | 396 |
| Residual Deviance | 953.5 |
| AIC | 983.5 |
| LR Chi$^2$ | 48.7 |
| Pr(>Chi2) | <0.01 |
| *Note:* | *p<0.1; **p<0.05; ***p<0.01 |

**Table 2.** We performed several proportional odds models to test whether our assumption of continuous response variables significantly impacted the model results. This model is comparable to Table 3 in the main body of the text. The outcome variable is willingness to pay. The significant variables and the direction of the effect is the same in both the proportional odds model shown below and the multiple regression shown in Table 3 of the main text.

| Independent Variable | Odds Ratio |
|---|---|
| Race | |
|     Latino | 1.15 |
|     Black /African American | 1.20 |
|     Asian or Pacific Islander | 0.94 |
|     Multi-racial | 0.75 |
|     Other | 0.58** |
| Gender (Female = 0) | 0.93 |
| Education | 1.09* |
| Age | 0.81*** |
| Flood zone | 1.03 |
| Time | 0.28*** |
| Exposure to ads | 0.99 |
| Concern for flooding | 1.32*** |
| Observations | 1115 |
| Residual Deviance | 3906.3 |
| AIC | 3944.3 |
| LR Chi$^2$ | 213.3 |
| Pr(>Chi2) | <0.01 |
| *Note:* | *p<0.1; **p<0.05; ***p<0.01 |

APPENDIX 4 – OLS PATH ANALYSIS WITH KNOWLEDGE VARIABLE

**Table 1**. OLS regression coefficients. This analysis utilizes the 2016 and 2017 datasets. Following equations (1) and (2) in the main body of the text, concern is the mediating variable ($M$) and willingness to pay ($WTP$) is the outcome variable ($Y$). Standard errors of the direct effect are presented in parentheses. Indirect effect coefficients ($a_i b$) are presented with the bootstrapped 95% confidence interval in brackets.

| Independent Variable (IV) | Direct Effects | | | | Indirect Effects |
|---|---|---|---|---|---|
| | IV → Concern | | IV → WTP | | IV → Concern → WTP |
| | $a$ | p | $c$ | p | $a_i b$ |
| Race (White = 0) | | | | | |
| Latino | 1.17 (0.25) | <0.01*** | -0.35 (0.33) | 0.28 | **0.22 [0.09, 0.37]** |
| Black/African American | 0.24 (0.16) | 0.13 | 0.13 (0.20) | 0.52 | 0.04 [-0.02, 0.12] |
| Asian or Pacific Islander | 0.18 (0.30) | 0.62 | 0.02 (0.38) | 0.95 | -0.03 [-0.10, 0.17] |
| Multi-racial | 0.31 (0.33) | 0.34 | -0.39 (0.42) | 0.35 | 0.06 [-0.10, 0.22] |
| Other | -0.05 (0.23) | 0.84 | -0.52 (0.29) | 0.07 | -0.01 [-0.10, 0.08] |
| Gender (Female = 0) | -0.29 (0.11) | <0.01*** | -0.03 (0.13) | 0.81 | **-0.05 [-0.11, -0.01]** |
| Education | -0.16 (0.05) | <0.01*** | 0.01 (0.07) | 0.87 | **-0.03 [-0.06, -0.01]** |
| Age | 0.08 (0.04) | 0.05** | -0.28 (0.05) | <0.01*** | |
| Flood zone | 0.35 (0.25) | 0.16 | 0.12 (0.32) | 0.70 | |
| Knowledge | 0.08 (0.07) | 0.28 | -- | | |
| Exposure to ads | 0.03 (0.03) | 0.45 | -- | | |
| Concern for flooding | -- | | 0.19 (0.05) | <0.01*** | |
| $R^2$ | 0.07 | | 0.08 | | |
| MSE | 1.95 | | 3.15 | | |
| F Statistic | 5.05*** (df = 11) | | 5.74*** (df = 10) | | |
| N | 720 | | 720 | | |

*Note:* *p<0.1; **p<0.05; ***p<0.01
**Bold** indicates that the confidence interval for the indirect estimate does not contain zero
Letters *a, b*, and *c* indicate coefficients displayed in Equations (1) and (2)